\documentclass[preprint,12pt]{elsarticle}
\usepackage{graphicx} %
\usepackage{color, xcolor, colortbl}
\usepackage[frozencache=true,cachedir=.]{minted}

\newcommand{\mr}[1]    {\textcolor{black}{#1}}
\newcommand{\minr}[1]    {\textcolor{black}{#1}}

\makeatletter
\let\myorg@bibitem\bibitem
\def\bibitem#1#2\par{%
  \@ifundefined{bibitem@#1}{%
    \myorg@bibitem{#1}#2\par
  }{%
    \begingroup
      \color{\csname bibitem@#1\endcsname}%
      \myorg@bibitem{#1}#2\par
    \endgroup
  }%
}
\makeatother

\newcounter{req}[section]
\newcommand{\req}[3]{\refstepcounter{req}\medskip\label{#3}\noindent\textbf{O\thereq~-~#1}:~#2 \rmfamily\medskip}

\date{}

\title{Qubernetes: Towards a Unified Cloud-Native Execution Platform for Hybrid Classic-Quantum Computing}

\begin{document}

\begin{frontmatter}
\author[inst1]{Vlad Stirbu}
\author[inst1]{Otso Kinanen}
\author[inst1]{Majid Haghparast}
\author[inst1]{Tommi Mikkonen}

\affiliation[inst1]{organization={University of Jyväskylä},%
            city={Jyväskylä},
            country={Finland}}

\begin{abstract}
\noindent\textbf{Context:} The emergence of quantum computing proposes a revolutionary paradigm that can radically transform numerous scientific and industrial application domains. The ability of quantum computers to scale computations beyond what the current computers are capable of implies better performance and efficiency for certain algorithmic tasks.

\noindent\textbf{Objective:} However, to benefit from such improvement, quantum computers must be integrated with existing software systems, a process that is not straightforward. In this paper, we propose a unified execution model that addresses the challenges that emerge from building hybrid classical-quantum applications at scale.

\noindent\textbf{Method:} \mr{Following the Design Science Research methodology}, we proposed a convention for mapping quantum resources and artifacts to Kubernetes concepts. Then, in an experimental Kubernetes cluster, we conducted experiments for scheduling and executing quantum tasks on both quantum simulators and hardware.

\noindent\textbf{Results:} The experimental results demonstrate that the proposed platform Qubernetes (or Kubernetes for quantum) exposes the quantum computation tasks and hardware capabilities following established cloud-native principles, allowing seamless integration into the larger Kubernetes ecosystem.

\noindent\textbf{Conclusion:} The quantum computing potential cannot be realised without seamless integration into classical computing. By validating that it is practical to execute quantum tasks in a Kubernetes infrastructure, we pave the way for leveraging the existing Kubernetes ecosystem as an enabler for hybrid classical-quantum computing. 

\end{abstract}

\begin{keyword}
quantum software \sep hybrid classical-quantum software \sep containers \sep \mr{quantum software development lifecycle \sep cloud-native computing}
\end{keyword}

\end{frontmatter}

\section{Introduction}

Quantum computers have demonstrated the potential to revolutionize various fields, including cryptography, drug discovery, materials science, and machine learning, by leveraging the principles of quantum mechanics. However, the current generation of quantum computers, known as noisy intermediate-scale quantum (NISQ) computers~\mr{\cite{nisq}}, suffer from noise and errors, making them challenging to operate. Additionally, the development of quantum algorithms requires specialized knowledge in the field of quantum mechanics and mathematics, which is not readily available to the majority of software professionals. These factors pose a significant entry barrier to leveraging the unique capabilities of quantum systems.

For the existing base of business applications, classical computing has already proven its capabilities across a diverse range of solutions. However, some of the computations they must perform can be accelerated with quantum computing, much like graphical processing units (GPUs) are used today. Therefore, quantum systems should not function in isolation, but they must coexist and interoperate with classical systems. 
To this end, the current way of building and operating quantum computers hinders their adoption, as application developers have to learn the bespoke way in which their programs are executed on the hardware. To make matters worse, the quantum simulator of the hardware target used for execution has to be explicitly selected, which blurs the line between the development and the operational phase in a product or software development lifecycle.

This paper proposes an approach where the focus is placed on the orchestration of classical and quantum computations. Kubernetes\footnote{https://kubernetes.io/}, a widely used system for automating deployment, scaling, and management of containerized applications, is used as the underlying infrastructure. In this approach, the quantum computations are packaged as containers that are executed on quantum-capable nodes alongside classical computations. Constructed in this way, Qubernetes -- the quantum-enhanced Kubernetes --- is tailored to fit hybrid classical-quantum applications.

 The rest of this paper is organized as follows. In Section \ref{sec:background}, we present the fundamental concepts of quantum computing, the quantum software development and key challenges faced by the developers and the hardware operators of hybrid classic-quantum systems. In Section \ref{sec:methodology}, we introduce the methodological background of this research. In Section \ref{sec:requirements}, we introduce the objectives of the solution, crystallised as requirements that need to be satisfied by a unified cloud-native hybrid classical-quantum computing execution platform. In Section \ref{sec:qubernetes}, we introduce Qubernetes, a Kubernetes platform extension that enables the execution of heterogeneous classic-quantum computing tasks. In Section \ref{sec:evaluation} we describe the experimental setup and the application scenarios used to validate the Qubernetes concept. In Section \ref{sec:discussion} we discuss how Qubernetes addresses the requirements and the needs of software developers. \minr{In Section \ref{sec:threats} we address the threats to validity}. Concluding remarks are provided in Section \ref{sec:conclusions}.

\section{Background \mr{and motivation}}
\label{sec:background}

\subsection{\mr{Quantum computing fundamentals}}

\mr{Qubits, which stands for quantum bits, are the fundamental units of quantum information in quantum computing. Unlike conventional bits, which can exist in one of two states (0 or 1), qubits can exist in multiple states simultaneously, thanks to the principles of superposition and entanglement, which are unique to quantum mechanics \cite{nielsen2010quantum}. %
This new computing paradigm enables the development of a new breed of algorithms \cite{Montanaro2016} that leverage the qubit capabilities to speed up the performance of computational tasks beyond what is possible with the existing classical computers \cite{GYONGYOSI}. For example, factoring large numbers using classical algorithms has exponential complexity, while using Shor's algorithm has polynomial complexity.}

\mr{The physical implementation of quantum computers can be split into two categories: specialized (e.g., special purpose computers designed to solve optimization problems using \textit{annealing} programming approach) or general-purpose (e.g., allowing programming of individual qubits using pulses or gate programming approaches). The current technological candidates for building gate-based general-purpose quantum computers fit within one of the following categories: \textit{superconducting} -- tiny superconducting materials are cooled to extremely low temperatures to manifest their quantum properties, \textit{trapped ion} -- ions are trapped within electromagnetic fields, or \textit{photonic} -- quantum information stored in photons can be manipulated and transmitted over long distances. In the longer term, the \textit{topological} quantum computers, leveraging the collective properties of ensembles of particles, will overcome the current NISQ limitations and achieve fault-tolerant operations \cite{gill}. Although these quantum computers are not yet advanced enough to achieve fault-tolerance or reach the scale required for quantum advantage \cite{sandersquantum, zhuquantum}, they provide an experimentation platform to develop new generations of hardware and quantum algorithms and validate quantum technology in real-world use cases. Whether a quantum computer is general-purpose or specialized, the selection of quantum qubit implementation technology can enhance hardware efficiency for specific problem classes \cite{qubitlayout, qubitperformance}. To use the hardware effectively, application developers must consider these differences when designing and optimizing the software's functionality and operations.}

Further, the concept of distributed quantum computers~\cite{distributed-quantum-computing}, which interlink multiple distinct quantum machines through quantum communication networks, emerges as a potential solution to amplify the available quantum volume~\cite{quantum-volume}, beyond what is possible using a single quantum computer. Nevertheless, the intricacies inherent in the distributed quantum computers remain hidden from users, as compilers aware of the distributed architecture of the target system shield them from such complexities. In essence, the quantum compiler plays a vital role in achieving the effective execution of generic quantum circuits on existing physical hardware platforms, making the compilers an active research area in quantum computing~\cite{haghparast2023quantum}.

\subsection{\mr{Quantum development kits}}

\begin{figure}[t]
    \centering
    \includegraphics[width=0.7\textwidth]{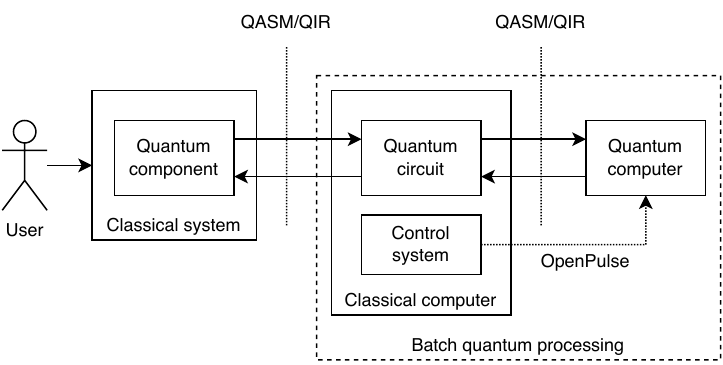}
    \caption{Quantum computing model: components and interfaces}
    \label{fig:computing-model}
\end{figure}

A typical hybrid classic-quantum software system is understood as a classical program that has one or more software components that are implemented using quantum technology, as depicted in Figure~\ref{fig:computing-model}. A quantum component relies on quantum algorithms~\cite{Montanaro2016}, which are transformed into quantum circuits. The quantum circuit describes quantum computations in a machine-independent language, such as quantum assembly (QASM)~\cite{openqasm}. This circuit is translated by a computer that controls the quantum computer in a machine-specific circuit and a sequence of operations, such as pulses~\cite{openpulse}, that control the operation on individual hardware qubits. The translation process, implemented using quantum compilers, encompasses supplementary actions like breaking down quantum gates, optimizing quantum circuits, and providing fault-tolerant iterations of the circuit.

\mr{Application developer use tools like Qiskit\footnote{https://qiskit.org} and Cirq\footnote{https://quantumai.google/cirq} for writing, manipulating and optimizing quantum circuits. These Python libraries allow researchers and application developers to interact with nowadays' NISQ computers, allowing them to run quantum programs on a variety of simulators and hardware designs, abstracting away the complexities of low-level operations and allowing researchers and developers to focus on algorithm design and optimization.}

\mr{Tools like TensorFlow Quantum\footnote{https://www.tensorflow.org/quantum} and PennyLane\footnote{https://pennylane.ai} play a crucial role in facilitating the development of machine learning quantum software. These frameworks provide high-level abstractions and interfaces that bridge the gap between quantum computing and classical machine learning. They allow researchers and developers to integrate quantum algorithms seamlessly into the machine learning development process by providing access to quantum simulators and hardware, as well as offering a range of quantum-friendly classical optimization techniques. TensorFlow Quantum leverages the power of Google's TensorFlow ecosystem, enabling the combination of classical and quantum neural networks for hybrid quantum-classical machine learning models. PennyLane offers a unified framework for developing quantum machine learning algorithms, supporting various quantum devices and seamlessly integrating them with classical machine learning libraries. These tools provide a foundation for researchers to explore and experiment with quantum machine learning, accelerating the progress and adoption of quantum computing in the field of machine learning.}

\subsection{\mr{Notebooks, simulators, and proxy access to quantum hardware}}

\mr{Jupyter\footnote{https://jupyter.org} notebooks and quantum simulators play a vital role in supporting developers of quantum programs. Jupyter provides an interactive and collaborative environment where developers can write, execute, and visualize their quantum code in an accessible manner. They allow for the combination of code, explanatory text, and visualizations, making it easier to experiment, iterate, and document the development process. Quantum simulators, on the other hand, enable developers to simulate the behavior of quantum systems without the need for physical quantum hardware. These simulators provide a valuable testing ground for verifying and debugging quantum algorithms, allowing developers to gain insights into their performance and behavior before running them on actual quantum devices. Developers can iterate quickly, gain a deeper understanding of quantum concepts, and refine their quantum programs efficiently.}

\mr{Traditional cloud computing providers, such as AWS Braket\footnote{https://aws.amazon.com/braket/}, Azure Quantum\footnote{https://learn.microsoft.com/en-us/azure/quantum/}, Google Quantum AI\footnote{https://quantumai.google} or IBM Quantum\footnote{https://quantum-computing.ibm.com}, offer comprehensive quantum development services. These services are designed to optimize the development process with integrated tools like Jupyter notebooks and task schedulers. Developers can create quantum applications and algorithms across multiple hardware platforms simultaneously. This approach ensures flexibility, allowing fine-tune algorithms for specific systems while maintaining the ability to develop applications that are compatible with various quantum hardware platforms.}

\subsection{\mr{Hybrid classical-quantum computing approaches}}

High-Performance Computing (HPC) is the mainstream approach for running scientific and engineering simulations at scale. Integrating the quantum computing and HPC software stacks enables quantum technology to accelerate parts of the simulations. Two notable approaches for integrating the two software stacks are HPC-QC~\cite{hpc-qc-linking}, which leverages the Open Message Passing Interface (OpenMPI\footnote{https://www.open-mpi.org}) compatible architectures, and XACC~\cite{XACC} approach based on the OSGi\footnote{https://www.osgi.org} architecture.

\mr{Similarly, the existing base of cloud applications can benefit from using quantum computing to accelerate the appropriate computational tasks, a trend that is not overlooked by the major quantum development toolkit providers. For example,} Qiskit's quantum-serverless~\cite{quantum-serverless} proposes a cloud-based approach for running hybrid classical-quantum programs. The proposed programming model, conforming to the RAY\footnote{https://www.ray.io} computing framework, makes it easy to scale Python workloads on a Kubernetes cluster in which the quantum execution environment is represented by a distributed Qiskit runtime that allows transparent access to multiple QPUs.

\subsection{\mr{Development process}}

\begin{figure}[t]
    \centering
    \includegraphics[width=\textwidth]{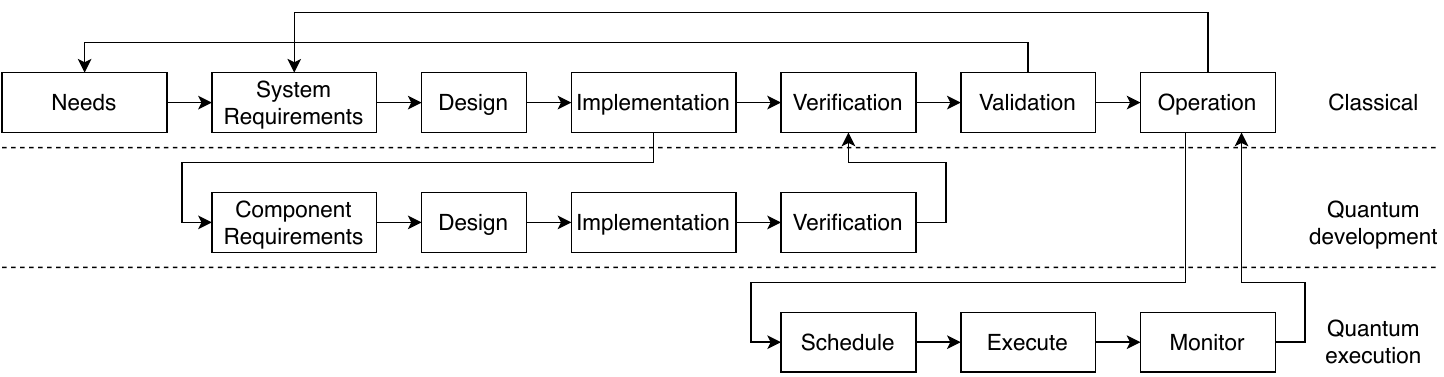}
    \caption{The software development lifecycle model for hybrid classical-quantum systems}
    \label{fig:sdlc}
\end{figure}

The software development life-cycle (SDLC) of hybrid classic-quantum applications consists of a multi-faceted approach \cite{stirbu2023fullstack}, as depicted in Figure~\ref{fig:sdlc}. At the top level, the classical software development process starts by identifying user needs and deriving them into system requirements. These requirements are transformed into a design and implemented. The result is verified against the requirements and validated against user needs. Once the software system enters the operational phase, any detected anomalies are used to identify potential new system requirements, if necessary. A dedicated track for quantum components is followed within the SDLC~\cite{sdlc}, specific to the implementation of quantum technology. The requirements for these components are converted into a design, which is subsequently implemented on classic computers, verified on simulators or real quantum hardware, and integrated into the larger software system. During the operational phase, the quantum software components are executed on actual quantum hardware. The scheduling ensures efficient utilization of the scarce quantum hardware resources, while monitoring capabilities enable the detection of anomalies throughout the operational stage.

\mr{As quantum computers are a scarce resource, it is not practical to develop quantum software components directly on hardware. Instead, developers can use simulators that use commonly available and less expensive classical resources (e.g., CPUs and GPUs) for the early stages of development and testing. As simulators become more sophisticated, being able to simulate the noise of actual hardware, developers can perform fast iterations with confidence. Only when the components are mature enough the development can be continued on actual the hardware that will be used during the execution phase. This approach ensures that the use of quantum resources is effective.}

\mr{Commercial entities, like QuantumPath \cite{qpath}, provide an integrated offering that covers multiple developments, including requirements management, editing and source code version control, and remote execution via proxy to quantum hardware. The integrated approach has near-term advantages as it lowers the entry barrier into a technologically complex environment. However, in the long term, as quantum technology is integrated into existing classical applications, the development methodologies and the tooling that support them will be inherited from what is already used for classical software development by the respective organizations. This is a particularly important concern for regulated industries (e.g., finance or medical~\cite{calmcompliance}) where regulatory-related automation is implemented in tools like JIRA\footnote{https://www.atlassian.com/software/jira}/Polarion\footnote{https://polarion.plm.automation.siemens.com} -- project and requirements management, and GitHub\footnote{https://github.com}/GitLab\footnote{https://about.gitlab.com} -- version control and code level change management.}

\subsection{\minr{Towards cloud-native quantum computing}}

\minr{Quantum technology has the ability to deliver quantum advantage for an array of applications (e.g.,  machine learning~\cite{quantumml} or optimizations~\cite{cloudnativeapps}), that can be implemented in cloud-native environments. When used within this context, the quantum technology must be properly integrated into the larger technological ecosystem (e.g. Kubernetes), and using modern DevOps practices~\cite{ebertdevops}, leveraging containers as the standard way of packaging software artifacts, and a high degree of automation employed at every stage of the SDLC.}

\begin{figure}[t]
    \centering
    \includegraphics[width=\textwidth]{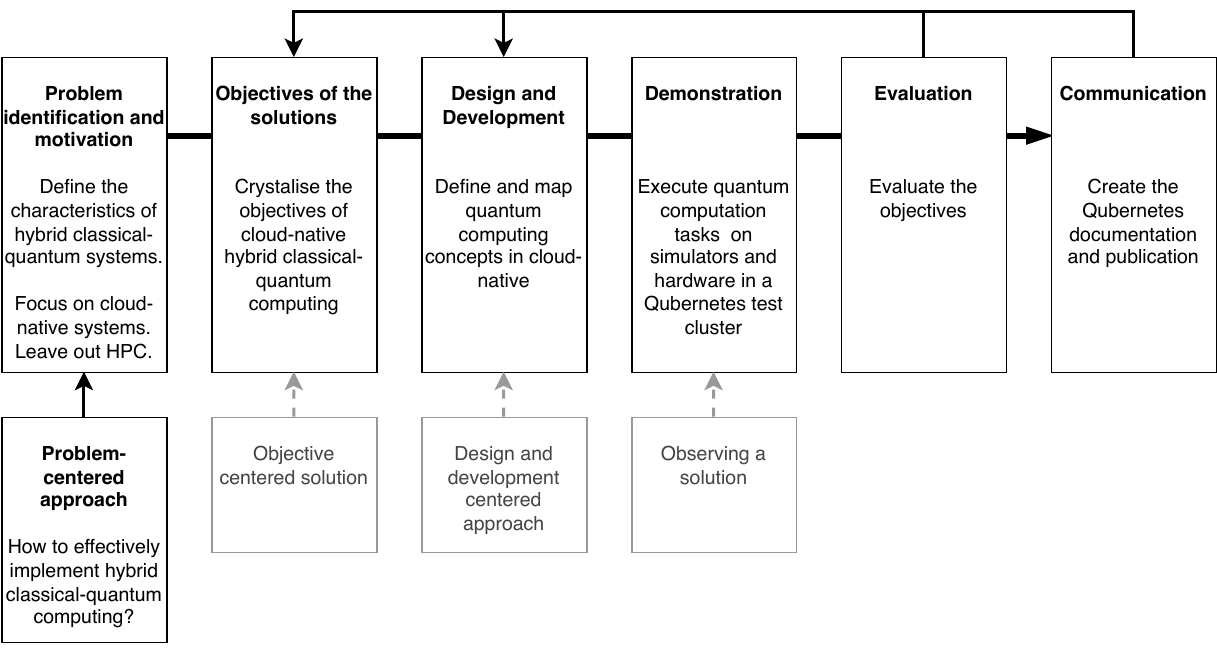}
    \caption{\mr{Design science research methodology applied to Qubernetes development}}
    \label{fig:sdrm-q8s}
\end{figure}

\section{\mr{Methodology}}
\label{sec:methodology}

\mr{The Qubernetes concept was developed using the problem-centric approach of the Design Science Research Methodology (DSRM) \cite{dsrm}. The starting point was to answer the research question \textit{How to effectively implement hybrid classical-quantum computing?} \minr{The research question was translated into a set of six objectives that need to be met by the solution to enable cloud-native integration. Further, in the \textit{Design and Development} phase we introduced how quantum computing concepts like quantum computer and computation tasks are exposed in to Kubernetes as quantum nodes and jobs. Then, for the \textit{Demonstration} phase, we described a test cluster, conforming to the Qubernetes convention, and provided an example quantum jobs developed using the Qiskit toolkit that is executed on all targets: CPU and GPU simulators (e.g., Qiskit-Aer), and quantum hardware (e.g., HELMI). For the \textit{Evaluation} phase, we have discussed how the Qubernetes solution addresses the objectives.} The process is depicted in Figure~\ref{fig:sdrm-q8s}.}

\minr{We have carefully considered the reproducibility of the test environment and opted for an approach in which the essential software artifacts are included in the paper using the established conventions for each technology: YAML specifications for serialized Kubernetes objects\footnote{https://kubernetes.io/docs/concepts/overview/working-with-objects/}, Dockerfile for the container descriptions\footnote{https://docs.docker.com/reference/dockerfile/}, and Python source for the simple quantum test program developed using Qiskit toolkit. The steps describing setting up a Kubernetes cluster, configuring the internal container registry, build an publish container images to registry or interacting with the cluster using the \texttt{kubectl}\footnote{https://kubernetes.io/docs/reference/kubectl/}, have been omitted for brevity, as they are covered by ample documentation on the respective projects' websites. Nevertheless, we have provided throughout the manuscript, whenever necessary, footnotes with the links that lead to the relevant online documentation. We acknowledge that our access to the HELMI quantum computer is attributed to our university’s membership in the consortia that owns the hardware, a circumstance that isn’t easily replicable. However, the CPU and GPU capabilities of the test cluster can be replicated by anyone with access to general-purpose computing and an Nvidia-compatible GPU, which are commercially available off-the-shelf products. We believe that this approach strikes the right balance between completeness and brevity, allowing the reader not only to replicate our results but to continue experimentation.}

\section{\mr{Objectives}}
\label{sec:requirements}

\mr{The shift to cloud computing has simplified the process of developing scalable applications. However, to fully harness the benefits of cloud computing, applications must adhere to cloud-native architectural principles~\cite{cloudnativearch}. This entails designing applications as small, loosely coupled components that can be bundled with their dependencies into portable containers and deployed on the immutable infrastructure. By leveraging the service discovery, load-balancing, and self-healing capabilities inherent in cloud platforms, development teams, comprising both software development and operations expertise, can automate the software development lifecycle and streamline delivery processes. Furthermore, emphasizing observability through integrated monitoring and logging offers valuable insights into performance, health, and behavior, empowering teams to swiftly respond to potential anomalies.}

\begin{figure}[t]
    \centering
    \includegraphics[width=0.55\textwidth]{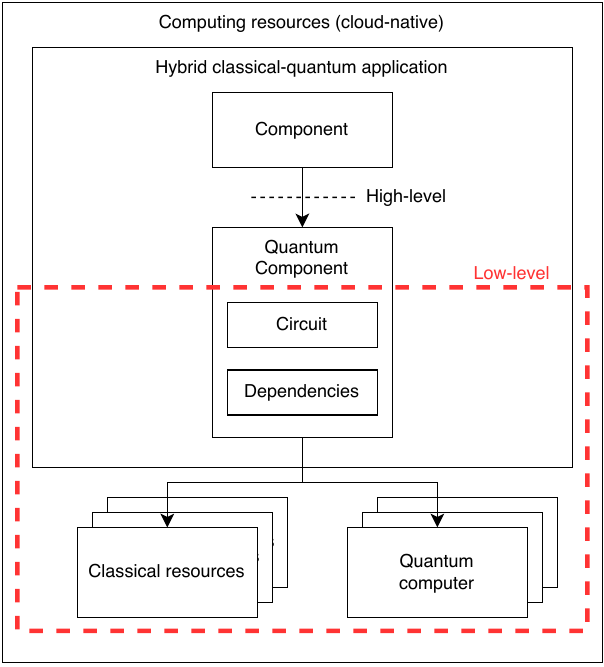}
    \caption{\mr{The solution boundaries within the hybrid classical-quantum application domain}}
    \label{fig:scope}
\end{figure}

Kubernetes is the industry-standard container orchestration platform for automating deployment, scaling, and management of containerized cloud-native applications. Developed as an open-source solution by Cloud Native Computing Foundation (CNCF)\footnote{https://www.cncf.io/}, \mr{together with the myriad of projects that offer supporting functionality, it allows users to deploy applications on the managed infrastructure of the major cloud providers (e.g., AWS EKS\footnote{https://aws.amazon.com/eks/}, Azure AKS\footnote{https://azure.microsoft.com/en-us/products/kubernetes-service}, or GCP GKE\footnote{https://cloud.google.com/kubernetes-engine}), smaller or regional cloud providers, or on-prem -- using own infrastructure. The reach functionality and wide industry adoption make Kubernetes the prime candidate for developing a cloud-native execution platform for hybrid classical-quantum computing.}

\mr{Quantum computing technology holds the potential to enhance the performance of cloud applications, particularly in domains such as machine learning and optimizations~\cite{quantumoptimization}. To facilitate seamless integration, the implementation of quantum components should align with existing development conventions and practices established in classical applications whenever possible. It's crucial to acknowledge that cloud-native applications are developed using a diverse array of programming languages and frameworks. In the realm of machine learning alone, there are various tools such as KubeFlow\footnote{https://www.kubeflow.org}, Seldon Core\footnote{https://www.seldon.io/solutions/seldon-core}, and RAY, to name a few. Consequently, a cloud-native solution for exposing quantum computing resources needs to focus on the \textit{low-level} interface between containerized workloads and simulators/hardware. Simultaneously, it should maintain an open \textit{high-level} interface between the classical and the quantum components, allowing for flexibility and interworking with different programming languages and frameworks, as illustrated in Figure \ref{fig:scope}. The following objectives crystallize the focus on the low-level interface described above.}

\begin{figure}[t]
    \centering
    \includegraphics[width=0.6\textwidth]{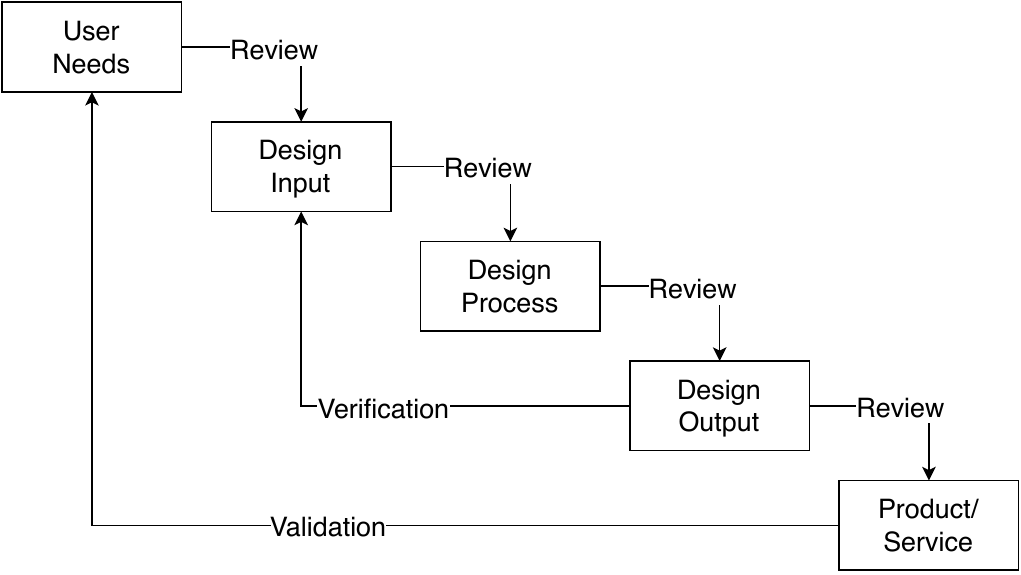}
    \caption{Design controls}
    \label{fig:design-control}
\end{figure}

\req{Design control and SDLC}{The design controls are part of a comprehensive quality system that covers the lifetime of a product or service. The process ensures that the user needs are met by the resulting product or service and that the design inputs and outputs on which the design process is based are verified through a rigorous review process, see Figure~\ref{fig:design-control}. They are based upon established quality assurance and engineering principles~\cite{iso9001}, covering changes to the product, service, or manufacturing process design, including those occurring long after a device has been introduced to the market. From a quantum software perspective, the software component developed using quantum technology needs to be validated and packaged in a format that is appropriate for execution during the quantum execution phase.}{req:design-control}

\req{Runtime support}{
The quantum programming frameworks (e.g., Qiskit or Cirq) employ distinct methods for exposing the quantum hardware as backends. As the framework includes a runtime for running the code, they are responsible for converting the input circuits, which are machine-independent, into machine-specific configurations using an internal representation expressed in QASM. Alternatively, an open and extensible toolchain and runtime based on intermediate representations for quantum programs that extend the LLVM compiler framework~\cite{llvm} are currently under development in the QIR Alliance\footnote{https://www.qir-alliance.org}. The QIR compiler has the ability not only to convert between the machine-independent and the machine-dependent circuits but also to mix intermediate representations originating from different quantum programming languages expressed as QIR. Further, the QIR ecosystem enables developers to create programs with complex classical and quantum instructions via its interoperability with LLVM.  These aspects of the execution environment have to be exposed at the platform level so that users can execute their quantum software on the appropriate hardware.
}{req:runtime}

\req{Programming model}{
Gate-level and pulse-level quantum programming are two distinct approaches used to control and manipulate quantum computers. In gate-level programming, quantum operations are expressed as a sequence of quantum gates that act on qubits. These gates are akin to logic gates in classical computing and are specified in a quantum circuit. Gate-level programming provides a high-level, hardware-independent representation of quantum algorithms. Most quantum programming frameworks support gate-level programming, e.g., Qiskit, Cirq, or TKET\footnote{https://www.quantinuum.com/developers/tket}. Similarly, machine learning-oriented quantum programming (e.g., Pennylane\footnote{https://pennylane.ai}) are gate-based~\cite{quantum-programming-languages}. On the other hand, pulse-level quantum programming involves direct manipulation of the microwave or laser pulses that drive the qubits. This level of programming is hardware-centric and enables fine-grained control over the quantum operations, providing opportunities for optimizing quantum algorithms. Pulse-level programming is well-suited for practitioners who want to harness the full potential of quantum hardware via specialized programming languages (e.g., Jaqal\footnote{https://www.sandia.gov/quantum/quantum-information-sciences/projects/qscout-jaqal/}, Qiskit Pulse\footnote{https://qiskit.org/documentation/apidoc/pulse.html}, or SimuQ~\cite{peng2023simuq}).
}{req:programming}

\req{Scheduling}{The scheduler is a software component that has the responsibility to find the appropriate resources required for executing correctly a quantum software component. Besides the basic functionality, the scheduler might consider additional inputs that affect its decisions. For example, the energy requirements for completing the job vs the cost of the energy can play a significant role in deciding the time when to schedule the execution. Similarly, from a time perspective, the scheduler can do more than act as a queue so that quantum executions that need to be completed fast are prioritized first, while the others are scheduled when the quantum hardware utilization decreases.}{req:scheduling}

\req{Execution}{The execution is the phase during which the quantum software component is run on the actual hardware. The execution typically involves the preparation of the hardware, a step performed by the control software that runs on a classical computer. Following the preparation, the quantum program is executed a number of times, with the results being collected and aggregated into a data structure that includes a probability distribution of the results.}{req:execution}

\req{Monitoring}
{The monitoring component performs comprehensive observation of the system performance targeted to the \textit{users} and to the \textit{operators} of the platform. Monitoring the execution allows the users to determine if there are anomalies in the execution that can lead to modification of the program. Similarly, monitoring allows operators to determine how the quantum hardware is utilized and detect how to improve resource utilization. Monitoring also fulfills the enabling layer of billing.}{req:monitoring}

\section{\mr{Qubernetes: design and concepts}}
\label{sec:qubernetes}

Qubernetes (Q8s) is a quantum computing-aware extension of Kubernetes. \mr{In this section, we describe how the quantum computing resources are mapped to the Kubernetes native concepts, serving as the foundation for building cloud-native hybrid classical-quantum applications.} %

\subsection{Quantum resource mapping overview}

\begin{figure}[t]
    \centering
    \includegraphics[width=0.9\textwidth]{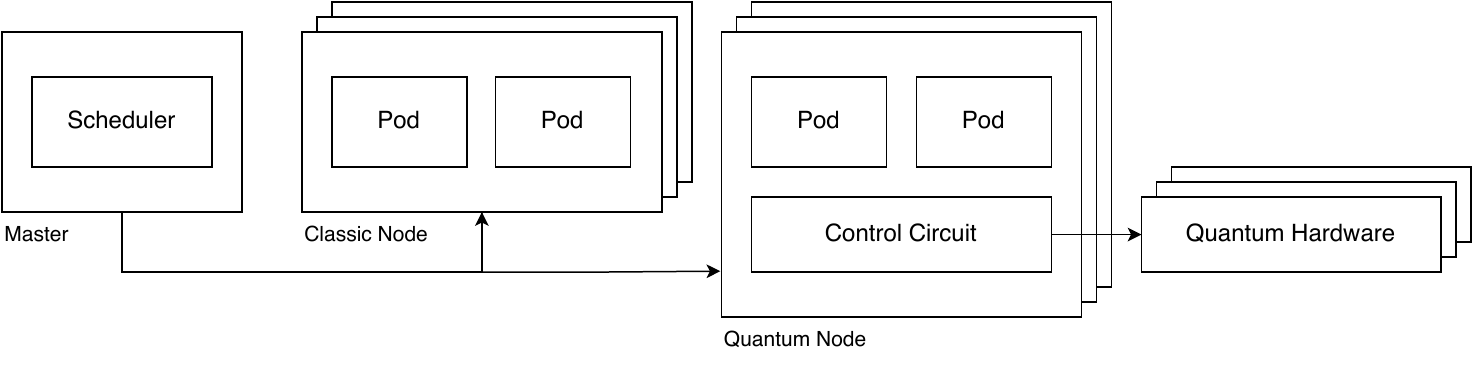}
    \caption{Qubernetes: quantum aware Kubernetes}
    \label{fig:qubernetes}
\end{figure}

In contrast to traditional Kubernetes, Q8s introduces the following pivotal additions: \mr{the quantum-capable node definition} and the quantum job \mr{definition that facilitates execution of quantum computations} on quantum-capable nodes. Quantum nodes seamlessly integrate quantum hardware and its associated control circuit capabilities into the Kubernetes cluster, while the quantum-aware scheduler is able to schedule jobs that instantiate the pods that need access to quantum hardware on the corresponding quantum nodes, as depicted in Figure \ref{fig:qubernetes}.

\subsection{Quantum node}

The quantum capable node joining the cluster is identified using specific labels (e.g., \texttt{accelerator}), and the QPU's capacity in their \texttt{Node} specification (e.g., \texttt{vendor.example.com/qpu}), see Listing \ref{listing:node}. The capacity indicated by the node is used by the scheduler to allocate pods on compatible nodes. As current quantum hardware is typically able to execute one task at a time, the value 1 means that the node is able to execute a task, while the value 0 indicates that it is not.

\begin{listing}[h]
\begin{minted}[fontsize=\footnotesize,linenos=false,highlightlines={5,8}]{yaml}
apiVersion: v1
kind: Node
metadata:
  labels:
    accelerator: qpu
status:
  capacity:
    vendor.example.com/qpu: 1
\end{minted}
\caption{Quantum computing capable node specification}
\label{listing:node}
\end{listing}

\subsection{Quantum job}

A \textit{Job} in Kubernetes is a workload resource designed to spawn a single Pod and ensure its reliable execution until completion. Given that quantum programs typically adhere to a batch execution model, reusing the Job workload is a well-suited choice.

\begin{listing}[h]
\begin{minted}[fontsize=\footnotesize,linenos=true,highlightlines={6-18}]{yaml}
apiVersion: batch/v1
kind: Job
metadata:
  name: quantum-job
spec:
  template:
    spec:
      nodeSelector:
        accelerator: qpu
      containers:
      - name: quantum-task
        image: registry.example.com/program:v1.2.3
        command: ["./extrypoint.sh"]
        resources:
          requests:
            vendor.example.com/qpu: 1
          limits:
            vendor.example.com/qpu: 1
\end{minted}
\caption{Quantum job specification}
\label{listing:job}
\end{listing}

The specific quantum task that needs to be executed as part of the Job is described by the \texttt{spec.template} key that includes a cue for the scheduler that the pod needs to be executed on a quantum capable node (e.g., \texttt{nodeSelector}), and it needs one slice of the specific hardware capacity (e.g., \texttt{vendor.example.com/qpu}).

\begin{listing}[h]
\begin{minted}[fontsize=\footnotesize,linenos=true,highlightlines={}]{yaml}
apiVersion: v1
kind: Pod
metadata:
  name: quantum-pod
spec:
  nodeSelector:
    accelerator: qpu
  containers:
    - name: quantum-task
      image: "registry.example.com/program:v1.2.3"
      resources:
        requests:
          vendor.example.com/qpu: 1
        limits:
          vendor.example.com/qpu: 1
\end{minted}
\caption{Pod specification created from the template described in the Job}
\label{listing:pod}
\end{listing}

\subsection{Scheduling and execution}

Kubernetes has sophisticated scheduling capabilities for classical computing that are able to handle heterogeneous computing capabilities like CPUs with different architectures (e.g., amd64 or arm64), GPUs (e.g., AMD, Intel, Nvidia), or even more exotic accelerators like TPUs (e.g., on Google Kubernetes Engine) or FPGAs. Using the labels and capabilities exposed by the quantum capable nodes, and the node selection preferences and the computing needs requested by pods, the default scheduler (e.g., \texttt{kube-scheduler}), without being aware of quantum computing internals, can create the pods, move them in \texttt{Pending} state, and wait till the appropriate nodes become available.

Once scheduled, a Pod moves into \texttt{Running} state, during which the quantum circuit is actually executed on the quantum hardware. Once the execution ends successfully, the pod state changes to \texttt{Succeeded}, and the corresponding Job becomes \texttt{Completed}. In case the execution fails, the pod status changes to \texttt{Failed}. The Job output can be fetched using \texttt{kubectl logs jobs/quantum-job}, as for any Kubernetes jobs.

\subsection{Logging and monitoring}

Logging is the process of capturing, storing, and analyzing the data generated by containers, applications, and infrastructure within a Kubernetes cluster. It plays a crucial role in monitoring, troubleshooting, and maintaining the health and performance of containerized applications and the underlying infrastructure. Kubernetes logging typically involves the collection of log data from various sources, such as containers, pods, and nodes, and centralizing it for analysis and visualization. Effective logging at the quantum node and pod level helps Kubernetes administrators and developers gain valuable insights into the application's behavior, diagnose issues, and ensure the reliability and security of the containerized quantum workloads.

Monitoring is an essential aspect of managing containerized applications within Kubernetes clusters. It involves the continuous collection, analysis, and visualization of data related to the performance, health, and resource utilization of both the applications and the underlying infrastructure. Kubernetes monitoring provides real-time insights into the behavior of containers, pods, nodes, and other resources, enabling administrators to proactively identify and resolve issues, optimize resource allocation, and ensure the reliability and scalability of the entire environment. Administrators can leverage tools such as Prometheus\mr{\footnote{https://prometheus.io}}, Grafana\mr{\footnote{https://grafana.com}}, or other Kubernetes-native monitoring solutions to enable operators to gain a comprehensive understanding of the cluster's operational status by tracking metrics, setting up alerts, or creating detailed dashboards. This data-driven approach is fundamental for maintaining the availability and performance of applications in dynamic, containerized environments.

\section{\mr{Demonstration}}
\label{sec:evaluation}

This section \mr{describes the environment used to demonstrate the use} of the Qubernetes platform. We start with a description of the experimental cluster in which the \mr{demonstration} was conducted. Then we describe the scenarios used for running quantum programs inside the test Qubernetes cluster.

\subsection{Experimental cluster setup}

\begin{figure}[t]
  \centering
  \includegraphics[width=\textwidth]{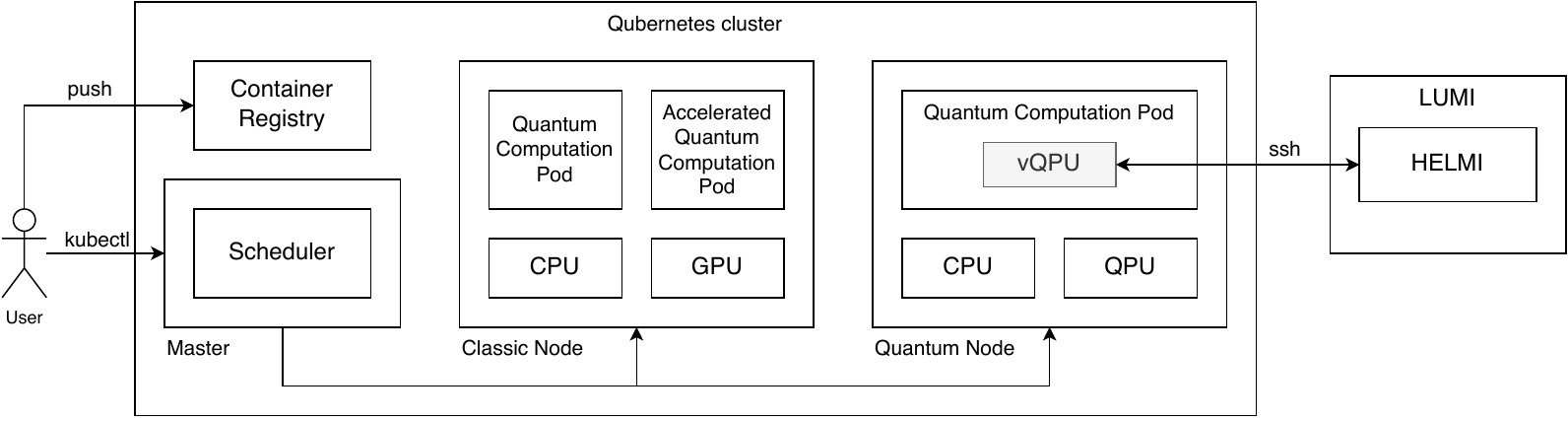}
  \caption{Experimental \mr{Qubernetes cluster} setup}
  \label{fig:experimental-setup}
\end{figure}

The evaluation of Qubernetes was performed on a Kubernetes cluster containing both classical and quantum computing resources (see Figure. \ref{fig:experimental-setup}). The classical nodes had CPU and GPU capabilities, \mr{allowing quantum computations to be executed in simulators, including the ones supported by Nvidia's cuQuantum\footnote{https://developer.nvidia.com/cuquantum-sdk}}. The quantum node exposed the QPU functionality as a \mr{a \textit{virtual QPU}}, implemented by a classical program \mr{(e.g., the \texttt{entrypoint.sh} script included in the container)} that sends commands over secure shell (ssh) to the IQM 5-qubit computer attached to the LUMI supercomputer operated by CSC\footnote{https://docs.csc.fi/computing/quantum-computing/overview/} in Finland.

\begin{listing}[!h]
\begin{minted}[fontsize=\footnotesize,linenos=true,highlightlines={}]{python}
from qiskit import QuantumCircuit, transpile
from qiskit_aer import AerSimulator

# Use Aer's AerSimulator
simulator = AerSimulator()

# Create a Quantum Circuit acting on the q register
circuit = QuantumCircuit(2, 2)

# Add a H gate on qubit 0
circuit.h(0)

# Add a CX (CNOT) gate on control qubit 0 and target qubit 1
circuit.cx(0, 1)

# Map the quantum measurement to the classical bits
circuit.measure([0, 1], [0, 1])

# Compile the circuit for the support instruction set (basis_gates)
# and topology (coupling_map) of the backend
compiled_circuit = transpile(circuit, simulator)

# Execute the circuit on the aer simulator
job = simulator.run(compiled_circuit, shots=shotsAmount)

# Grab results from the job
result = job.result()

# Returns counts
counts = result.get_counts(compiled_circuit)
print("\nTotal count for 00 and 11 are:", counts)
\end{minted}
\caption{Simplified test program intended to run on CPU}
\label{listing:program}
\end{listing}

\begin{figure}[t]
  \centering
  \includegraphics[width=0.5\textwidth]{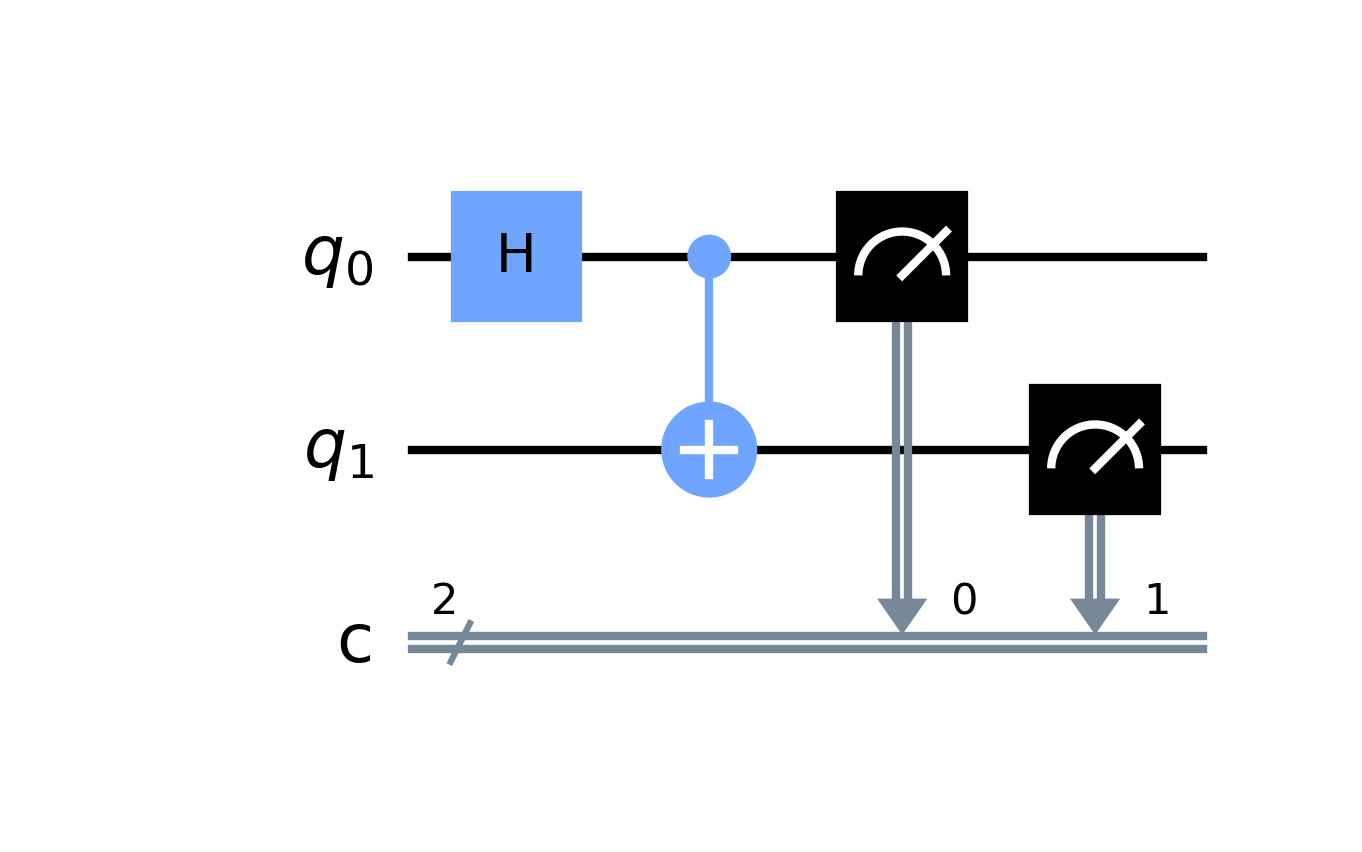}
  \caption{\mr{The representation of the quantum circuit used in the experiment}}
  \label{fig:circuit}
\end{figure}

The test application was a simple quantum program developed using the Qiskit framework, depicted in Listing \ref{listing:program}. \mr{The program contains all the structural elements expected in a typical quantum program regardless of the programming framework used (e.g., Cirq, PennyLine, etc.): backend selection (line 5), quantum circuit definition (lines 8-17), transpilation of the machine-independent circuit to the backend-specific circuit (line 21), execution on the backend (line 24), and using the results (lines 27-31). The simple quantum circuit consisting of two qubits and a 2-qubit gate (depicted in Figure~\ref{fig:circuit}) is light enough in terms of gate complexity that can be executed in all target environments (e.g., CPU or GPU-based simulators or actual quantum computers), but still demonstrates a measurable result of a quantum computation task.}  %

The program is packaged as a container\mr{, together with the appropriate dependencies and the \texttt{entrypoint.sh} script,} then published to the cluster's internal \mr{container} registry. \mr{The blueprint of the container specification is presented in Listing \ref{listing:container}}.%

\begin{listing}[h]
\begin{minted}[fontsize=\footnotesize,linenos=true,highlightlines={}]{docker}
FROM --platform=amd64 nvidia/cuda:11.6.2-base-ubuntu20.04

COPY requirements.txt .
RUN pip install - r requirements.txt

COPY test.py .
COPY entrypoint.sh .

CMD ["./entrypoint.sh"]
\end{minted}
\caption{\mr{The container blueprint for executing the quantum task in a Pod}}
\label{listing:container}
\end{listing}

The program is executed in the cluster as a Job that requires the execution completion of one Pod following the Kubernetes conventions. The quantum jobs are submitted, and the results of the execution are fetched using \texttt{kubectl} commands \texttt{apply} and \texttt{logs}, as expected in a Kubernetes cluster.

\subsection{Execute the quantum \mr{computation} task in simulator}

The experiment's objective is to run a test program on a classical node within the cluster, utilizing the high-performance quantum computing simulator \texttt{qiskit-aer}\footnote{https://github.com/Qiskit/qiskit-aer}, which includes realistic noise models. Initially, the program is executed on a node that solely relies on CPU resources, as evident in the Job specification by the absence of resource requests (e.g., as seen in lines 14-18 in Listing \ref{listing:job}).

Subsequently, the program is adapted to employ \texttt{qiskit-aer-gpu}, the GPU-accelerated version of the simulator. %
This modified execution takes place on a GPU-enabled node within the cluster, as indicated by the necessary hardware specified in the Job configuration (e.g., lines 16 and 18 in Listing \ref{listing:job} are altered to \texttt{nvidia.com/gpu:~1}).

\subsection{Execute the quantum \mr{computation} task on quantum hardware}

The aim of the experiment is to run the test program on the HELMI quantum computer. The test program is adjusted to utilize the HELMI backend\footnote{https://docs.csc.fi/computing/quantum-computing/helmi/running-on-helmi/}. An \mr{\texttt{entrypoint.sh}} script that communicates with HELMI via SSH, executes the required commands, and waits for their completion is added to the container image. The Job submission is scheduled to run on a designated node configured as described in Listing~\ref{listing:node}. \mr{The Job description has additional configuration that exposes the needed ssh keys in the running Pod, enabling \texttt{entrypoint.sh} script to communicate securely with HELMI.}

\section{Discussion}
\label{sec:discussion}

In this section, we first discuss how Qubernetes meets the \mr{objectives} for a hybrid classical-quantum cloud native execution platform. Additionally, we compare how Qubernetes compares with alternative approaches\mr{, and propose future research directions}.

\subsection{QPU-capable node implementation}

Within the experimental setup, the role of the quantum computer is assumed by the HELMI computer, operated by CSC. Our approach involves accessing the HELMI computer and executing the necessary commands to run the quantum program through an SSH session. Given that HELMI is an older system, this method of integrating its functionality into the Kubernetes cluster serves as a proof of concept. Fortunately, recent developments in quantum computing have seen new hardware vendors and cloud providers offering remote APIs for their quantum computers (e.g., Atos QML\footnote{https://pypi.org/project/qlmaas/} or AWS Braket). Further, ongoing research initiatives like \mr{European High-Performance Computing Joint Undertaking\footnote{https://eurohpc-ju.europa.eu/} (EuroHPC JU)} are working on defining \mr{\textit{Universal Quantum Access}~\cite
{eurohpc}}, a concept that would \mr{not only} enable access to various \mr{local and remote} quantum computers \mr{via standardised interfaces and protocols}, \mr{but would also facilitate the effective use of these quantum resources}. These advancements will facilitate a more straightforward implementation of quantum resources at the node level. Overall, Kubernetes has the ability to expose the runtime and hardware capabilities using node labels, fulfilling the \mr{intent of objectives} O\ref{req:runtime} and O\ref{req:programming}, \mr{and collect the logs entries from the Pods to a centralised drain (e.g., Prometheus), enabling monitoring according to objective O\ref{req:monitoring}.}

\subsection{Scheduling quantum tasks}

Currently, Kubernetes has native capabilities for basic scheduling, being able to execute quantum tasks on classic nodes using simulators and on quantum-capable nodes using actual hardware. Using the Job metaphor enables the equivalent experience as running the program using traditional means directly on hardware. However, the abstraction layers are low-level and might not be appropriate for advanced usage. More sophisticated scheduling mechanisms existing within the Kubernetes ecosystem can be used, e.g., Kueue\footnote{https://kueue.sigs.k8s.io}. Overall, the Kubernetes native scheduler and its extension points are able to fulfill the \mr{objective} O\ref{req:scheduling}.

\subsection{Quantum task execution unit}

Packaging quantum software components as containers (e.g., \mr{objective} O\ref{req:design-control}) provides an immutable artifact that enables Qubernetes to execute in a consistent and repeatable fashion the respective components. The container executed as a Pod within the context of completed Jobs corresponds to a unit of work that can be easily understood for monitoring purposes (e.g., \mr{objective} O\ref{req:monitoring}), and also for billing if needed. The approach decouples the execution from the design artifact. The selection of the appropriate execution environment moves to the quantum operational phase, implemented via scheduling. In the experimental setup, we have separate Job specifications for each node configuration (e.g., CPU, GPU, QPU). We can leverage configuration management mechanisms like Kustomize\footnote{https://kustomize.io} or Helm\footnote{https://helm.sh} to derive the specifications from one template that serves as a single source of truth.

\subsection{Quantum task abstraction level}

The quantum tasks are executed in Qubernetes using the Pod and Job-native objects. Although using \texttt{kubectl} to submit and execute these \textit{run once to completion} jobs is effective \mr{to demonstrate the low-level interface}, its usability might not be appropriate in all cases. We plan to use \mr{Kubernetes' native} higher-level concepts like \texttt{Service}\mr{\footnote{https://kubernetes.io/docs/concepts/services-networking/service/}} to investigate how to enable more sophisticated functionality, such as repetitive jobs triggered on demand. Though running tasks in batch mode (e.g., either as Jobs or Services) produces the expected outcomes, the timing can be non-deterministic, contingent on the quantum hardware's load. To address this, rather than relying on synchronous interaction mechanisms (e.g., run to completion for short jobs or request-response for services), we intend to investigate the potential of asynchronous approaches utilizing message queues (e.g., KubeMQ\footnote{https://kubemq.io}) or enterprise service bus (ESB) integration patterns to deliver a reactive experience.

\mr{The Service abstraction would be able to handle inputs/outputs, thus integrating the quantum components into the classical applications. However, classical cloud-native applications are developed using a wide range of programming languages and frameworks, and their components communicate using multiple protocols. Therefore, on the high-level interface, rather than imposing a unified integration approach, it is more beneficial to leverage the interaction patterns and protocols already used in the classical application context.}

\subsection{Kubernetes cluster management}

The initial approach was to rely on a partition (e.g., namespace in Kubernetes) of a managed cluster (e.g., Rahti\footnote{https://rahti.csc.fi} operated by CSC in Finland), where we had several nodes with CPUs and one node having GPU access. Due to the difficulties of adding the node that exposed the QPU resource to the cluster, we reverted to running our own cluster. Quantum hardware providers may find it necessary to manage Kubernetes clusters themselves to oversee node management, but they can effectively utilize the namespace feature to create isolated environments for accommodating multiple users simultaneously.

\subsection{Related approaches \mr{ and future research directions}}
\label{sec:related-approaches}

\mr{For quantum technology to truly deliver its potential and have a meaningful impact, it must seamlessly integrate into applications seeking a computational performance advantage. Proxy access solutions to remote hardware represent an initial step aimed at reducing entry barriers for developers wishing to run their programs on diverse hardware platforms. While effective for experimentation, these solutions have limitations in terms of scalability and fail to address the broader challenges of integrating and managing quantum software components within larger systems. Nonetheless, the functionality provided by proxy access solutions can serve as a foundation for implementing QPU-capable quantum nodes within a Qubernetes cluster, thereby granting users access to a more extensive array of quantum hardware resources.}

\mr{Comprehensive solutions such as QuantumPath, which span multiple SDLC phases, should adhere to the established practices governing the development of applications into which quantum software components are integrated. Given the existing fragmentation within the landscape, not only in terms of high-level approaches (such as HPC or cloud-native), a less opinionated and modular approach is preferred. This approach entails seamlessly folding quantum technology and its development practices into the existing framework of classical computing, ensuring compatibility and flexibility across various environments and methodologies.}

\mr{Similarly, even when cloud-native by design, frameworks like} quantum-serverless \cite{quantum-serverless}, which is highly optimised for Qiskit\mr{, has a narrow applicability as it imposes its own programming model. In comparison,} 
Qubernetes provides a generic execution engine that maps quantum tasks closer to the Kubernetes native concepts. As a result, our approach provides more flexibility, allowing non-Qiskit tasks to be executed on the same underlying quantum resources while event being able to expose parts of quantum-serverless functionality (e.g., the Qiskit multi-QPU runtime).

\mr{Looking forward, while Qubernetes demonstrates that it serves as a solid foundation to enable the introduction of quantum technology within the context of cloud-native applications, there are some areas that are still brittle and require further development. First, implementing the access from a quantum node to the backend requires bespoke solutions. We see the Quantum Universal Access activity as a key enabler for the effective use of quantum hardware also within cloud-native classical-quantum applications. Secondly, as the exiting quantum development kits (e.g., Qiskit, Cirq) have a monolithic architecture and cannot be easily combined, the developer experience using them together is poor. A compiler toolkit based on quantum intermediate representations (QIR) can improve the reuse and composability of software components developed with different frameworks. Finally,} the more sophisticated orchestration, monitoring capabilities, and integrations of the cloud-native computing~\cite{hpc-vs-k8s} \mr{have been identified as gaps by the HPC community. Their response was to establish the High Performance Software Foundation (HPSF) that aims to develop solutions that are aligned with Cloud Native Computing Foundation (CNCF), the home of cloud-native development. We expect that in the long term the technical implementations of the HPC and cloud-native computing to be much closer aligned than they are today.}

\section{\minr{Threats to validity}}
\label{sec:threats}

\minr{The threats to the validity of our study are discussed following to the categorization provided by Wholin et al.~\cite{wohlin}, in the context of applied research.}

\subsection{\minr{Internal threats}}

\minr{An internal threat to our study validity arises from developing the demonstrators only using the Qiskit toolkit. The mitigation in this case is that other popular Python toolkits (e.g., PennyLane or Cirq) have a similar software architecture that abstracts the hardware implementation regardless the target is a real quantum computer or a simulator implemented in CPU or GPU (leveraging the CUDA and cuQuantum toolkits). Further, the Kubeflow MPI Operator\footnote{https://github.com/kubeflow/mpi-operator} demonstrates that it is possible to run distributes tasks that typically require HPC-like infrastructure in Kubernetes, allowing a wider range of quantum state-vector device simulators (e.g., Pennylane Lightning Kokkos\footnote{https://docs.pennylane.ai/projects/lightning/en/stable/lightning\_kokkos/device.html}).}

\minr{Another internal threat to validity is the use of only one quantum computer (e.g., HELMI) to conduct execution experiments in our study. Considering that most quantum computers nowadays have bespoke ways to expose their functionality that is mapped typically to a backend in popular QDKs, we are forced to work with what the manufacturers provide. The emergence of standardised APIs (e.g., Universal Quantum Access) will enable consistent and uniform implementations of quantum nodes in Qubernetes clusters.}

\subsection{\minr{External threats}}

\minr{A threat to the external validity of our study is that exposing quantum computers as nodes in the Qubernetes cluster relies on adapting the bespoke solutions developed by their manufacturers or operators, which requires their cooperation. The mitigation of this threat is that as the industry is moving towards standardised APIs (e.g., Universal Quantum Access), it will become increasingly easy to integrate and use quantum computers in new environments without relying on manufacturers direct support.}

\subsection{\minr{Construct threats}}

\minr{A threat to construct validity is the long term viability of the Kubernetes Job as the primary mean to execute quantum computational tasks. The mitigation for this threat was to focus the study on the low-level interface between the containerized quantum workloads and the simulators/hardware. This approach allows the quantum pods to be reused into existing higher-level Kubernetes concepts (e.g., Services), or even develop completely new quantum-specific concepts using Custom Resource Definitions\footnote{https://kubernetes.io/docs/concepts/extend-kubernetes/api-extension/custom-resources}.}

\section{Conclusions}
\label{sec:conclusions}

Qubernetes demonstrates that Kubernetes has the capabilities that enable the development of hybrid classic-quantum at scale. Kubernetes already has the proper abstractions to enable both the utilization of quantum hardware and the execution of quantum software components along the classic software. We discussed the challenges that emerge from developing hybrid classical-quantum computers and proposed a hybrid architecture model building on a unified application-level view of software. %

\section*{Acknowledgements}
This work has been supported by the Academy of Finland (project DEQSE 349945) and Business Finland (project TORQS 8582/31/2022).

\end{document}